\documentclass{Interspeech2024}

\usepackage{hyperref}

\interspeechcameraready

\title{Enhancing Dysarthric Speech Recognition for Unseen Speakers via Prototype-Based Adaptation}

\name[affiliation={1}]{Shiyao}{Wang}
\name[affiliation={1}]{Shiwan}{Zhao}
\name[affiliation={1}]{Jiaming}{Zhou}
\name[affiliation={1}]{Aobo}{Kong}
\name[affiliation={1,*}]{Yong}{Qin}

\address{
  $^1$Nankai University, China 
  }
\email{wangshiyao@mail.nankai.edu.cn, zhaosw@gmail.com, zhoujiaming@mail.nankai.edu.cn, kongaobo9@163.com, 
qinyong@nankai.edu.cn }


\begin{document}

\maketitle

\renewcommand{\thefootnote}{}
\footnotetext{* Corresponding author.}

\begin{abstract}
    
Dysarthric speech recognition (DSR) presents a formidable challenge due to inherent inter-speaker variability, leading to severe performance degradation when applying DSR models to new dysarthric speakers. Traditional speaker adaptation methodologies typically involve fine-tuning models for each speaker, but this strategy is cost-prohibitive and inconvenient for disabled users, requiring substantial data collection. To address this issue, we introduce a prototype-based approach that markedly improves DSR performance for unseen dysarthric speakers without additional fine-tuning. Our method employs a feature extractor trained with HuBERT to produce per-word prototypes that encapsulate the characteristics of previously unseen speakers. These prototypes serve as the basis for classification. Additionally, we incorporate supervised contrastive learning to refine feature extraction. By enhancing representation quality, we further improve DSR performance, enabling effective personalized DSR.
We release our code at https://github.com/NKU-HLT/PB-DSR. 
\end{abstract}

\section{Introduction}
Dysarthria, a speech disorder caused by various factors such as neuropathy, muscle paralysis affecting speech, decreased muscle contractility, or motor incoordination, is frequently associated with conditions like cerebral palsy, Parkinson's disease, and head trauma. For those affected, the limited ability to use keyboards or touchscreens makes speech the most convenient means of interacting with devices like smartphones and smart home devices. However, dysarthria-induced changes in breathing, resonance, pronunciation, and prosody significantly impair the performance of speech recognition systems trained on typical speech \cite{34,27}. Addressing these impairments, dysarthric speech recognition (DSR) technology seeks to bridge the gap for dysarthric speakers, enabling seamless interaction with digital devices. While some studies \cite{18, 19, 15, 16} have focused on developing speaker-independent (SI) DSR models, the inherent variability among dysarthric speakers—due to differences in etiology, age, gender, speaking style, and severity of dysarthria—poses a significant challenge. This variability makes each speaker's speech patterns distinct, leading to notable performance degradation when SI models are applied to unseen dysarthric speakers.

Consequently, several studies \cite{14,33,9,20,2} have explored fine-tuning speech recognition models with data from the target speaker for personalized DSR. For instance, Shor et al. \cite{14} achieved this by fine-tuning a conventional speech recognition model with data from a specific dysarthric speaker, selectively fine-tuning only a subset of the network layers to avoid overfitting. Takashima et al. \cite{33} introduced a two-stage fine-tuning approach, initially leveraging data from multiple dysarthric speakers to develop an SI model that captures general dysarthric speech patterns, followed by further fine-tuning with target speaker data. Recently, Shahamiri et al. \cite{2} devised a novel DSR system that learns to recognize the word shapes spoken by dysarthric speakers and maps them to words. This system achieved optimal performance through the use of target speaker data adaptation. However, these personalized fine-tuning methods require extensive speech data from the target speaker and incur significant training costs. Moreover, the varying severity of dysarthria over time \cite{11} can compromise the long-term efficacy of personalized DSR models, necessitating continuous data collection and model optimization. This not only escalates training expenses but also poses substantial challenges for dysarthric speakers.

To address the aforementioned challenges, we introduce a prototype-based DSR (PB-DSR) method that leverages the unique pronunciation error characteristics of dysarthric speakers. Unlike traditional approaches that necessitate extensive model fine-tuning for each new speaker, our method requires only a minimal dataset to effectively adapt to individual speech patterns. Pronunciation errors, such as phoneme deletion, substitution, insertion, and distortion, are consistent within speakers. Previous studies \cite{28,29} attempted to address these errors by creating adaptive pronunciation dictionaries for each speaker, based on the analysis of error frequencies. However, these solutions were limited to the most common errors and did not adequately address the misrecognition of severe breath sounds and background noise as phonemes. In contrast, our approach targets word-level pronunciation errors, treating the speech for each word by an individual as a unique class that requires similarity rather than identical matches. Inspired by the principles of prototypical networks for few-shot learning \cite{3}, our PB-DSR method utilizes a feature extractor to create per-word prototypes from only few-shot samples, enabling rapid adaptation to the unique speech patterns of each speaker. We utilize the pre-trained HuBERT \cite{7} model, renowned for its general speech recognition capabilities, to extract speech features. To further refine its accuracy for dysarthric speech, HuBERT is initially fine-tuned with a specialized dysarthric speech dataset. Moreover, we integrate supervised contrastive learning (SCL) \cite{10} to enhance feature extraction, adopting a strategy from contrastive learning \cite{23} that optimizes feature representations by minimizing intra-class distances while maximizing inter-class separations, a technique proven effective in various speech tasks \cite{22,32,31,35}.

\begin{figure*}[th]
  \centering
  \includegraphics[width=\linewidth]{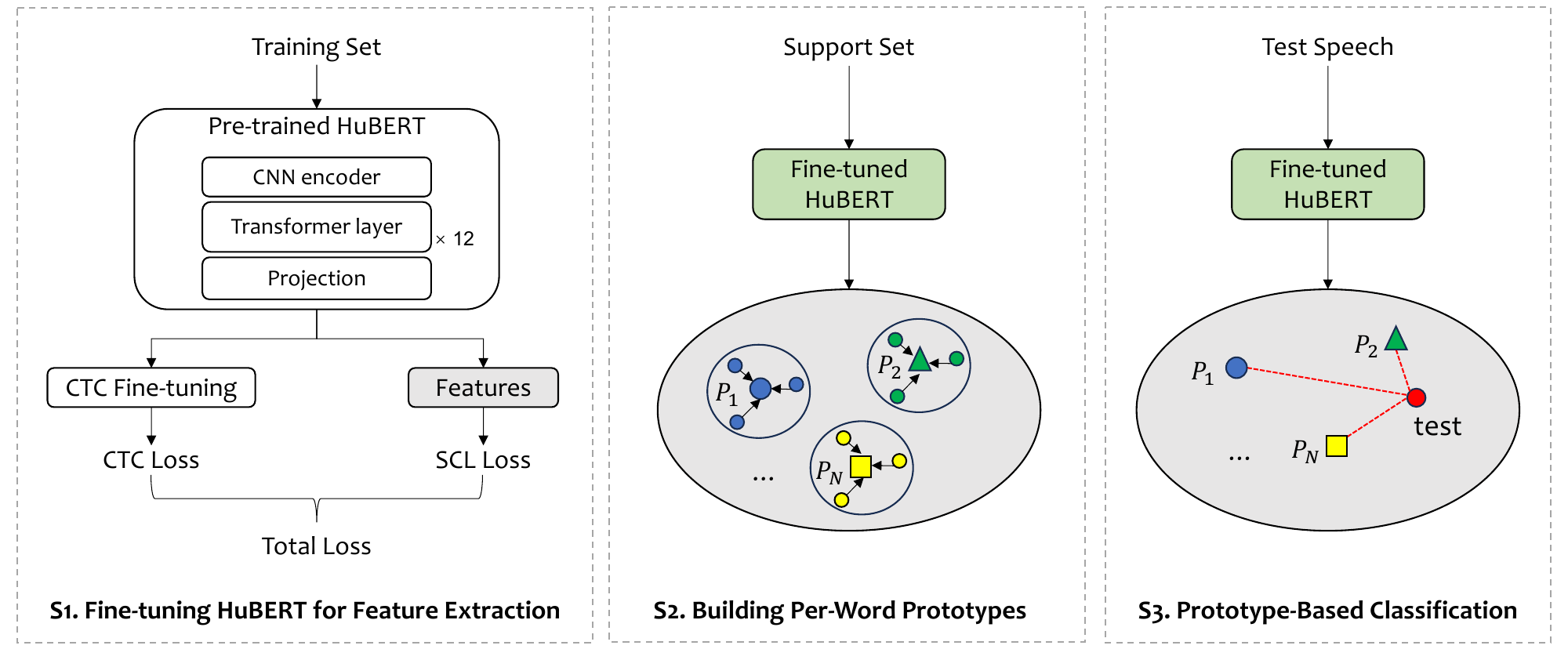}
  \caption{Prototype-based DSR comprises three stages: fine-tuning HuBERT for feature extraction, building per-word prototypes, and prototype-based classification.}
  \label{fig:f1}
  \vspace{-10pt} 
\end{figure*}

To empirically validate the efficacy of our PB-DSR method, we conduct experiments with the UASpeech dataset \cite{5}. The results demonstrate notable improvements in speech recognition for unseen dysarthric speakers. Specifically, the PB-DSR method achieves an average absolute reduction in Word Error Rate (WER) of 15.59\% compared to  its Speaker-Independent (SI) counterpart. Furthermore, incorporating SCL loss into our DSR model training to refine feature extraction leads to an additional 1.21\% reduction in WER.

The main contributions of this work are as follows:

\begin{itemize}    
    \item We propose a prototype-based DSR approach that offers a rapid and effective method for improving the recognition of speech from unseen dysarthric speakers. 
    \item We combine CTC loss with SCL loss to train the DSR model to improve performance by learning better feature representations.
    \item We have demonstrated the effectiveness of the proposed methods on the UASpeech dataset.
\end{itemize}

\section{Proposed Methods}
\subsection{Prototype-Based DSR} 
We present an overview of prototype-based DSR (PB-DSR) illustrated in Figure \ref{fig:f1}, encompassing three key stages: fine-tuning HuBERT for feature extraction, building per-word prototypes, and prototype-based classification.

\noindent\textbf{Stage 1. Fine-tuning HuBERT for Feature Extraction}

We fine-tune the pre-trained HuBERT on the DSR task, employing both CTC loss and SCL loss to effectively guide the model's training during the fine-tuning phase. The fine-tuned HuBERT serves as the feature extractor.

\noindent\textbf{Stage 2. Building Per-Word Prototypes}

We classify the speech of different words spoken by the unseen dysarthric speaker as distinct categories, and utilize the limited amount of speech data provided by the unseen dysarthric speaker as a support set. We employ the fine-tuned HuBERT to extract the features of speech from the support set. Subsequently, we average the features of the same word to build per-word prototypes. These prototypes serve as representations of the respective words in the DSR.

\noindent\textbf{Stage 3. Prototype-Based Classification}

We employ the fine-tuned HuBERT to extract the feature of test speech. It is important to emphasize that both the test speech and the speech in the support set are from the same unseen dysarthric speaker. Following this, we compute the distance between the test speech feature and each prototype. Ultimately, we choose the word represented by the nearest prototype as the recognition result.

\subsection{Combining CTC Loss with SCL Loss}
The CTC loss is widely utilized in speech recognition task due to its ability to accommodate data with potential mismatches in alignment between text and speech. Considering the presence of pauses and heavy breathing sounds in dysarthric speech, we opt for the CTC loss function as the primary loss function.

Furthermore, SI models perform poorly in recognizing the speech from speakers with more severe dysarthria, or speech from unseen dysarthric speakers, possibly attributed to the model's challenge in distinguishing the speech of different words in these speakers. Hence, we integrate the CTC loss with the SCL loss as the total loss function, with the goal of enhancing the model's ability to learn improved feature representations by simultaneously augmenting inter-class distances and diminishing intra-class distances.

In SCL, within a training batch, samples sharing the identical label are considered positive samples for each other, while those with distinct labels are treated as negative samples. Consequently, the SCL loss, denoted as $L_{SCL}$, is formulated as follows:
\begin{align}
  L_{SCL} &=  \sum_{i \in I } \frac{-1}{|P(i)|} \sum_{p \in P(i) } log \frac{exp(x_i · x_p)/\tau}{\sum_{a \in A(i)  }exp(x_i · x_p)/\tau},
  \label{equation:eq2}
\end{align}
where $i \in I = \{1, . . . , N\}$ denotes the index of a speech, $A(i)$ denotes all indices except $i$ , and $x_{i}$ denotes feature extracted by HuBERT. $P(i)$ denotes all indices of the positive samples of sample $i$. The features representing the same word as sample $i$ are positive samples. $\tau$ is the temperature hyperparameter.

The total loss, denoted as $L_{Total}$, combines the CTC loss $L_{CTC}$ with the SCL loss $L_{SCL}$ as follows:
\begin{align}
  L_{Total}&= L_{CTC} + L_{SCL}.
  \label{equation:eq1}
\end{align}

\section{Experiments}
\subsection{Dataset}
We assess the efficacy of the proposed method using the UASpeech dataset \cite{5}, comprising speech data from 15 dysarthric speakers and 13 control speakers. Details of intelligibility for 15 dysarthric speakers are presented in Table \ref{tab:Dysarthric speaker information}, with intelligibility denoting the percentage of speech comprehended by the average listener \cite{26}. The dataset is partitioned into 3 blocks (block 1, 2, and 3), comprising a total of 455 isolated words. Each speaker reads 10 digits, 26 radio alphabet words, 19 computer commands, and 100 common words three times. Each block encompasses one set of these speech. Additionally, there are 300 uncommon words, with each word read once by each speaker. These uncommon words are distributed across the blocks, with each block containing 100 uncommon words. The UASpeech data is recorded with an 8-microphone array, where microphone 1 is utilized for tone synchronization, and microphones 2 to 8 capture the speech signals. Due to data volume requirements for fine-tuning, the process utilizes data from 7 microphones. To expedite PB-DSR and assess its few-shot learning capabilities, per-word prototypes are exclusively built using data solely from microphone 5 (M5).

Below are descriptions of data splits:
\begin{itemize}
     \item \textbf{TRAIN}: All speech from control speakers and speech from block 1 and block 3 of dysarthric speakers.
     \item \textbf{CTEST}: The speech from block 2 of all dysarthric speakers excluding uncommon words speech.
\end{itemize}

\subsection{Experimental Settings}
\renewcommand{\thefootnote}{\arabic{footnote}} 
\subsubsection{The Settings of DSR Models}
We fine-tune HuBERT using the UASpeech dataset under the Fairseq open-source framework\footnote{\url{https://github.com/facebookresearch/fairseq/blob/main/examples/hubert/README.md}} to train the DSR model. The DSR model has a parameter size of 94.84M and is trained using a single NVIDIA GeForce RTX 3090. We employ the pre-trained HuBERT Base\footnote{\url{https://dl.fbaipublicfiles.com/hubert/hubert\_base\_ls960.pt}}
, which has been pre-trained on the Librispeech 960h dataset\footnote{\url{http://www.openslr.org/12}}. The DSR model is trained using the configuration file base\_10h.yaml\footnote{\url{https://github.com/facebookresearch/fairseq/blob/main/examples/hubert/config/finetune/base_10h.yaml}}, employing an initial learning rate of $10^{-5}$, 32000 warmup steps, and a batch size of 40. We set the parameter $\tau$ in Eq \ref{equation:eq2} to 0.07. The utilized dictionary comprises the 455 words from the UASpeech dataset, along with the tokens blank, $<s>$, $<pad>$, $</s>$, and $<unk>$. Training will cease if the loss fails to decrease for 10 consecutive epochs. The evaluation metric is the Word Error Rate (WER), measured in percentage.
{
}
\begin{table}[th]
    \caption{Dysarthric speaker intelligibility information}
    \label{tab:Dysarthric speaker information}
    \centering
    \resizebox{\columnwidth}{!}{ 
    \begin{tabular}{c c c}
        \toprule
        \textbf{Intelligibility Level} & \textbf{Speaker ID}  & \textbf{Speech Intelligibility (\%)} \\   
        \midrule
            & M08 & 93 \\
            & M09 & 86 \\
        High& M10 & 93 \\
            & F05 & 95 \\
            & M14 & 90 \\
        \hline
            & M05 & 58 \\	
        Mid & F04 & 62 \\
            & M11 & 62 \\
        \hline
            & M07 & 28 \\
        Low & F02 & 29 \\
            & M16 & 43 \\
        \hline
            & M01 & 15 \\
            & M04 & 2 \\
    Very low & F03 & 6 \\
            & M12 & 7 \\
         \bottomrule
    \end{tabular}
    }
  \vspace{-10pt} 
\end{table}

\begin{table*}[th]
    \caption{WER (\%) on different settings. ``SI" and ``SD" denote speaker-independent and speaker-dependent models, respectively. ``FT" denotes fine-tuning, and ``+" denotes adding SCL loss.}
    \label{tab:total table} 
    \centering
    \begin{tabular}{c | c c c | c c | c c | c c c c}
        \toprule
        \textbf{Intelligibility} & \multicolumn{3}{c|}{SI for Seen} &  \multicolumn{2}{c|}{SI for Unseen} &\multicolumn{2}{c|}{SD w/o FT} & \multicolumn{4}{c}{SD w/ FT}
        \\   
        \textbf{Level} & \textbf{\cite{19} }& \textbf{V} & \textbf{V+} & \textbf{R} & \textbf{R+} & \textbf{PB-DSR} & \textbf{PB-DSR+} & \textbf{\cite{20}} & \textbf{\cite{2}} & \textbf{FT-R} & \textbf{FT-R+PB-DSR}\\
        \midrule
        High & 6.40 & \textbf{2.35}	& 2.57 & 10.28 & 8.27 & 6.30 & \textbf{5.34} & 11.80 & 8.56 & \textbf{4.31} & 5.12 \\
         Mid & 14.60 & 6.01	& \textbf{4.97} & 26.57 & 25.27 & 15.40 & \textbf{14.57} & 34.67 & 30.75 & 5.53 & \textbf{4.89}\\
         Low & 18.90 & 7.91	& \textbf{7.23} & 50.58 & 45.41 & 23.21 & \textbf{22.63} & 26.67 & 26.75	& 7.85 & \textbf{6.27}\\
    Very Low & 61.50 & \textbf{32.08} & 32.11 & 84.05 & 83.22 & 64.19 & \textbf{61.72} & 59.50 & 66.35 & 39.01 & \textbf{37.67}\\
    \midrule
    Average & 25.35 & 12.09 & \textbf{11.72} & 42.87 & 40.54 & 27.28 & \textbf{26.07} & 33.16 & 33.10 & 14.18 & \textbf{13.49}\\
         \bottomrule
    \end{tabular}
    \vspace{-15pt} 
\end{table*}

We first train the vanilla DSR model (denoted as \textbf{V}) using the \textbf{TRAIN} split, to handle the seen dysarthric speakers. To validate the effectiveness of the proposed PB-DSR method in enhancing the speech recognition performance for unseen dysarthric speakers, we iteratively designate each dysarthric speaker from the UASpeech dataset as `unseen'. Subsequently, we remove the data of the designated speaker from the \textbf{TRAIN} split, resulting in modified splits denoted as \textbf{R-TRAIN}. From this process, we obtain 15 distinct \textbf{R-TRAIN}, each corresponding to one of the dysarthric speakers in the dataset. 
We train 15 models on \textbf{R-TRAIN} using only the CTC loss function; these models are collectively referred to as \textbf{R}. Additionally, we train another set of 15 models employing the proposed combined loss function, which are denoted as \textbf{R+}. Subsequently, we apply the prototype-based method to both \textbf{R} and \textbf{R+} models, resulting in our \textbf{PB-DSR} and \textbf{PB-DSR+} configurations, respectively.

To further compare our proposed PB-DSR method with the conventional fine-tuning approach, we proceed to fine-tune each \textbf{R} model using data from the corresponding unseen speaker. These fine-tuned models are denoted as \textbf{FT-R}. Following this, we implement the prototype-based method on \textbf {FT-R}, resulting in \textbf {FT-R+PB-DSR} configuration. This step allows us to assess the efficacy of the prototype-based method in improving speech recognition performance on fine-tuned models.

\subsubsection{The Settings of Prototype-Based DSR}
Due to the peak behavior in CTC models \cite{36,38}, coupled with the fact that our DSR model is configured with words as the dictionary, we observe that only the first frame is predicted as a word after the HuBERT feature passes through the CTC layer. Building on this observation, we compare the performance of PB-DSR when building prototypes using the full HuBERT features and solely relying on the first frame of HuBERT features. We find that utilizing only the first frame produces superior outcomes. Therefore, we utilize the first frame of the extracted HuBERT feature to build per-word prototypes.

We utilize the IndexFlatL2 index in the Faiss library \cite{25} to calculate the Euclidean distance between the feature of each test speech and per-word prototypes. 

\subsection{Results and Discussions}
All experimental results are detailed in Table \ref{tab:total table}. Initially, we benchmark the SI models for seen speakers, trained on the \textbf{TRAIN} split, against other studies to validate our model settings and the efficacy of the proposed loss function (Eq. \ref{equation:eq1}). Drawing from an understanding of dysarthric speech features, Bhat et al. \cite{19} devised static and dynamic data augmentation techniques for dysarthric speech, resulting in good speech recognition performance on the UASpeech dataset. As \cite{19} shares our data settings, i.e., evaluating model performance on the \textbf{CTEST} split, it serves as a direct comparison. Our models (\textbf{V} and \textbf{V+}) demonstrates a significant improvement in WER across all intelligibility levels compared to \cite{19}, with p-values \textless 0.05, highlighting the enhancements our approach brings to dysarthric speech recognition.

We then compare the DSR performance of \textbf{R} and \textbf{PB-DSR} for unseen dysarthric speakers. It becomes evident that the \textbf{PB-DSR} significantly enhances the DSR performance at each intelligibility level (p-value \textless 0.05). Specifically, it reduces the absolute values of WER by 3.98\% (high), 11.17\% (mid), 27.37\% (low), and 19.86\% (very low), respectively.
Subsequently, upon comparing \textbf{R} with \textbf{R+}, it can be observed that the addition of SCL loss enhances the speech recognition performance of DSR model for unseen speakers (p-value \textless 0.05). As the model learns better feature representation, the performance of \textbf{PB-DSR} is further enhanced, evidenced by a significant difference (p-value \textless 0.05) between \textbf{PB-DSR} and \textbf{PB-DSR+}.

We extend our comparison to include recent works by Shahamiri et al. \cite{20,2}, which similarly emphasize target speaker adaptation but with fine-tuning approaches. Notably, even without model fine-tuning, our approach achieves comparable performance to \cite{20,2} at the very low intelligibility level and outperforms them at other intelligibility levels. It's important to note, however, that these studies do not employ HuBERT, limiting direct comparisons under identical conditions. Consequently, we compare \textbf{PB-DSR} with \textbf{FT-R}, where \textbf{FT-R} involves fine-tuning. The performance of \textbf{PB-DSR} is relatively lower. Nevertheless, this outcome is deemed reasonable given that \textbf{PB-DSR} does not require any fine-tuning. Then we compare \textbf{FT-R+PB-DSR} with \textbf{FT-R}. We observe performance improvements in all intelligibility levels (p-value \textless 0.05) except for the high intelligibility level. This demonstrates that integrating the PB-DSR method with fine-tuned models further enhances DSR performance for unseen dysarthric speakers, underscoring the adaptability and effectiveness of the PB-DSR approach.

\subsection{Visualization}
\begin{figure}[th]
  \centering
  \includegraphics[width=\linewidth]{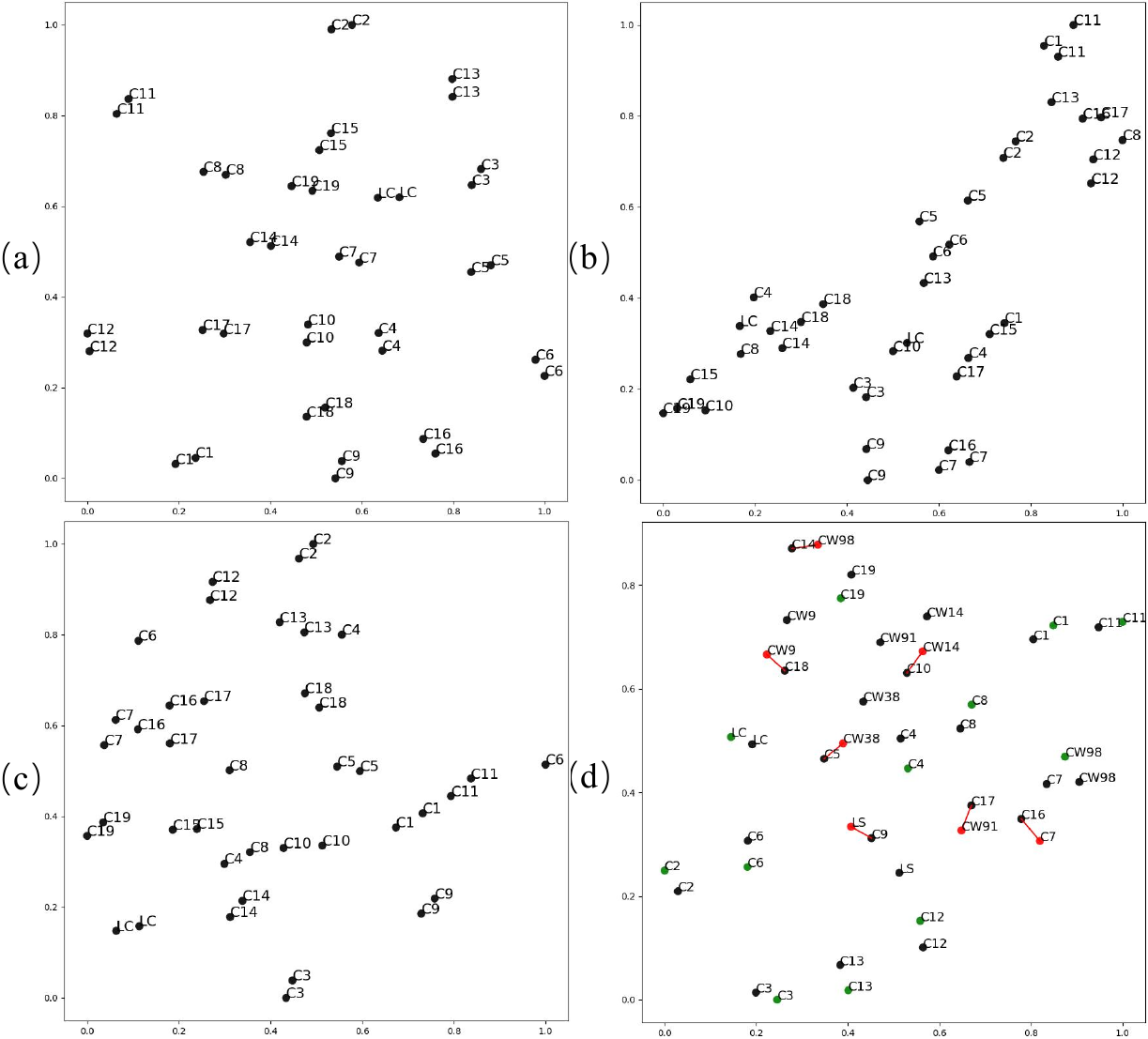}
\caption{Visualizations of speech feature distributions: (a) Seen speaker in \textbf{V}. (b) Unseen speaker in \textbf{R}. (c) Unseen speaker in \textbf{R+}. (d) Enhancing \textbf{R+} by \textbf{PB-DSR+}. Black dots in (d) represent prototypes, while green and red dots denote samples correctly and incorrectly classified by \textbf{R+}, respectively, with their labels from \textbf{R+} predictions. A red line connects a red dot to its correct prototype by \textbf{PB-DSR+}. Each label is the word ID in the UASpeech dataset.}   
  \label{fig:vis_total}
  \vspace{-18pt} 
\end{figure}
We illustrate the correspondence between DSR model performance and speech feature distribution by visualizing the first frame of HuBERT features from speech data. Figure \ref{fig:vis_total} (a) (b) (c) displays the feature distributions for the same speaker under different settings. For a seen speaker, the SI model \textbf{V} shows good speech recognition performance, reflected in the clustered pattern of speech features for the same word. Conversely, for unseen speakers, the SI model \textbf{R} exhibits poorer performance, evident in a more chaotic feature distribution. However, by integrating SCL, the \textbf{R+} model enhances the organization of speech feature distribution, demonstrating its potential to improve model performance across different speaker scenarios.

In Figure \ref{fig:vis_total} (d), we examine the speech recognition outcomes for an unseen speaker using \textbf{R+} and \textbf{PB-DSR+}. It's evident that features misrecognized by the \textbf{R+} model are nearly aligned with their correct prototypes, showcasing proximity but a crucial lack of precise 
recognition due to the model's inability to adjust to the novel speech patterns of the unseen speaker. Conversely, \textbf{PB-DSR+}, by employing straightforward distance metrics for classification, accurately recognizes these samples. This methodological refinement significantly boosts speech recognition accuracy for unseen speakers, highlighting the efficacy of PB-DSR in overcoming the adaptation challenges faced by conventional SI models.

\section{Conclusions}
We propose a prototype-based DSR method that effectively enhances the speech recognition performance for unseen dysarthric speakers without the need for fine-tuning or substantial data from those speakers. Additionally, we introduce supervised contrastive learning to enhance the performance of DSR. Comprehensive experiments on the UASpeech dataset affirm the effectiveness of our proposed method.


\section{Acknowledgments}
This work has been supported in part by  NSF China (Grant No.62271270).


\bibliographystyle{IEEEtran}
\bibliography{mybib}

\begin{thebibliography}{10}
\providecommand{\url}[1]{#1}
\csname url@samestyle\endcsname
\providecommand{\newblock}{\relax}
\providecommand{\bibinfo}[2]{#2}
\providecommand{\BIBentrySTDinterwordspacing}{\spaceskip=0pt\relax}
\providecommand{\BIBentryALTinterwordstretchfactor}{4}
\providecommand{\BIBentryALTinterwordspacing}{\spaceskip=\fontdimen2\font plus
\BIBentryALTinterwordstretchfactor\fontdimen3\font minus \fontdimen4\font\relax}
\providecommand{\BIBforeignlanguage}[2]{{%
\expandafter\ifx\csname l@#1\endcsname\relax
\typeout{** WARNING: IEEEtran.bst: No hyphenation pattern has been}%
\typeout{** loaded for the language `#1'. Using the pattern for}%
\typeout{** the default language instead.}%
\else
\language=\csname l@#1\endcsname
\fi
#2}}
\providecommand{\BIBdecl}{\relax}
\BIBdecl

\bibitem{34}
M.~Tu, A.~Wisler, V.~Berisha, and J.~M. Liss, ``The relationship between perceptual disturbances in dysarthric speech and automatic speech recognition performance,'' \emph{The Journal of the Acoustical Society of America}, vol. 140, no.~5, pp. EL416--EL422, 2016.

\bibitem{27}
L.~De~Russis and F.~Corno, ``On the impact of dysarthric speech on contemporary asr cloud platforms,'' \emph{Journal of Reliable Intelligent Environments}, vol.~5, pp. 163--172, 2019.

\bibitem{18}
\BIBentryALTinterwordspacing
A.~Hernandez, P.~A. P{\'e}rez-Toro, E.~N{\"o}th, J.~R. Orozco-Arroyave, A.~K. Maier, and S.~H. Yang, ``Cross-lingual self-supervised speech representations for improved dysarthric speech recognition,'' in \emph{Interspeech}, 2022. [Online]. Available: \url{https://api.semanticscholar.org/CorpusID:247940222}
\BIBentrySTDinterwordspacing

\bibitem{19}
C.~Bhat, A.~Panda, and H.~Strik, ``Improved asr performance for dysarthric speech using two-stage data augmentation,'' \emph{Proc. Interspeech 2022}, pp. 46--50, 2022.

\bibitem{15}
T.~Wang, S.~Hu, J.~Deng, Z.~Jin, M.~Geng, Y.~Wang, H.~Meng, and X.~Liu, ``{Hyper-parameter Adaptation of Conformer ASR Systems for Elderly and Dysarthric Speech Recognition},'' in \emph{Proc. INTERSPEECH 2023}, 2023, pp. 1733--1737.

\bibitem{16}
E.~Kim, Y.~Chae, J.~Sim, and K.~Lee, ``{Debiased Automatic Speech Recognition for Dysarthric Speech via Sample Reweighting with Sample Affinity Test},'' in \emph{Proc. INTERSPEECH 2023}, 2023, pp. 1508--1512.

\bibitem{14}
J.~Shor, D.~Emanuel, O.~Lang, O.~Tuval, M.~Brenner, J.~Cattiau, F.~Vieira, M.~McNally, T.~Charbonneau, M.~Nollstadt \emph{et~al.}, ``Personalizing asr for dysarthric and accented speech with limited data,'' \emph{arXiv preprint arXiv:1907.13511}, 2019.

\bibitem{33}
R.~Takashima, T.~Takiguchi, and Y.~Ariki, ``Two-step acoustic model adaptation for dysarthric speech recognition,'' in \emph{ICASSP 2020-2020 IEEE International Conference on Acoustics, Speech and Signal Processing (ICASSP)}.\hskip 1em plus 0.5em minus 0.4em\relax IEEE, 2020, pp. 6104--6108.

\bibitem{9}
D.~Wang, J.~Yu, X.~Wu, L.~Sun, X.~Liu, and H.~Meng, ``Improved end-to-end dysarthric speech recognition via meta-learning based model re-initialization,'' in \emph{2021 12th International Symposium on Chinese Spoken Language Processing (ISCSLP)}.\hskip 1em plus 0.5em minus 0.4em\relax IEEE, 2021, pp. 1--5.

\bibitem{20}
S.~R. Shahamiri, V.~Lal, and D.~Shah, ``Dysarthric speech transformer: A sequence-to-sequence dysarthric speech recognition system,'' \emph{IEEE Transactions on Neural Systems and Rehabilitation Engineering}, 2023.

\bibitem{2}
S.~R. Shahamiri, K.~Mandal, and S.~Sarkar, ``Dysarthric speech recognition using depthwise separable convolutions: Preliminary study,'' in \emph{2023 International Conference on Speech Technology and Human-Computer Dialogue (SpeD)}.\hskip 1em plus 0.5em minus 0.4em\relax IEEE, 2023, pp. 78--82.

\bibitem{11}
K.~Tomanek, K.~Seaver, P.-P. Jiang, R.~Cave, L.~Harrell, and J.~R. Green, ``An analysis of degenerating speech due to progressive dysarthria on asr performance,'' in \emph{ICASSP 2023-2023 IEEE International Conference on Acoustics, Speech and Signal Processing (ICASSP)}.\hskip 1em plus 0.5em minus 0.4em\relax IEEE, 2023, pp. 1--5.

\bibitem{28}
K.~T. Mengistu and F.~Rudzicz, ``Adapting acoustic and lexical models to dysarthric speech,'' in \emph{2011 IEEE International Conference on Acoustics, Speech and Signal Processing (ICASSP)}.\hskip 1em plus 0.5em minus 0.4em\relax IEEE, 2011, pp. 4924--4927.

\bibitem{29}
Y.~Sawa, R.~Takashima, and T.~Takiguchi, ``Adaptation of a pronunciation dictionary for dysarthric speech recognition,'' in \emph{2022 IEEE 4th Global Conference on Life Sciences and Technologies (LifeTech)}.\hskip 1em plus 0.5em minus 0.4em\relax IEEE, 2022, pp. 631--635.

\bibitem{3}
J.~Snell, K.~Swersky, and R.~Zemel, ``Prototypical networks for few-shot learning,'' \emph{Advances in neural information processing systems}, vol.~30, 2017.

\bibitem{7}
W.-N. Hsu, B.~Bolte, Y.-H.~H. Tsai, K.~Lakhotia, R.~Salakhutdinov, and A.~Mohamed, ``Hubert: Self-supervised speech representation learning by masked prediction of hidden units,'' \emph{IEEE/ACM Transactions on Audio, Speech, and Language Processing}, vol.~29, pp. 3451--3460, 2021.

\bibitem{10}
P.~Khosla, P.~Teterwak, C.~Wang, A.~Sarna, Y.~Tian, P.~Isola, A.~Maschinot, C.~Liu, and D.~Krishnan, ``Supervised contrastive learning,'' \emph{Advances in neural information processing systems}, vol.~33, pp. 18\,661--18\,673, 2020.

\bibitem{23}
T.~Chen, S.~Kornblith, M.~Norouzi, and G.~Hinton, ``A simple framework for contrastive learning of visual representations,'' in \emph{International conference on machine learning}.\hskip 1em plus 0.5em minus 0.4em\relax PMLR, 2020, pp. 1597--1607.

\bibitem{22}
L.~Wu, D.~Zong, S.~Sun, and J.~Zhao, ``A sequential contrastive learning framework for robust dysarthric speech recognition,'' in \emph{ICASSP 2021-2021 IEEE International Conference on Acoustics, Speech and Signal Processing (ICASSP)}.\hskip 1em plus 0.5em minus 0.4em\relax IEEE, 2021, pp. 7303--7307.

\bibitem{32}
X.~Wang, S.~Zhao, and Y.~Qin, ``{Supervised Contrastive Learning with Nearest Neighbor Search for Speech Emotion Recognition},'' in \emph{Proc. INTERSPEECH 2023}, 2023, pp. 1913--1917.

\bibitem{31}
J.~Zhou, S.~Zhao, N.~Jiang, G.~Zhao, and Y.~Qin, ``Madi: Inter-domain matching and intra-domain discrimination for cross-domain speech recognition,'' in \emph{ICASSP 2023-2023 IEEE International Conference on Acoustics, Speech and Signal Processing (ICASSP)}.\hskip 1em plus 0.5em minus 0.4em\relax IEEE, 2023, pp. 1--5.

\bibitem{35}
H.~Wang, X.~Zheng, and Y.~Qin, ``Intermediate-task learning with pretrained model for synthesized speech mos prediction,'' in \emph{2023 IEEE International Conference on Multimedia and Expo (ICME)}.\hskip 1em plus 0.5em minus 0.4em\relax IEEE, 2023, pp. 378--383.

\bibitem{5}
H.~Kim, M.~Hasegawa-Johnson, A.~Perlman, J.~Gunderson, T.~S. Huang, K.~Watkin, and S.~Frame, ``Dysarthric speech database for universal access research,'' in \emph{Ninth Annual Conference of the International Speech Communication Association}, 2008.

\bibitem{26}
H.~Chandrashekar, V.~Karjigi, and N.~Sreedevi, ``Investigation of different time-frequency representations for intelligibility assessment of dysarthric speech,'' \emph{Ieee transactions on neural systems and rehabilitation engineering}, vol.~28, no.~12, pp. 2880--2889, 2020.

\bibitem{36}
A.~Zeyer, R.~Schl{\"u}ter, and H.~Ney, ``Why does ctc result in peaky behavior?'' \emph{arXiv preprint arXiv:2105.14849}, 2021.

\bibitem{38}
J.~Zhou, S.~Zhao, Y.~Liu, W.~Zeng, Y.~Chen, and Y.~Qin, ``Knn-ctc: Enhancing asr via retrieval of ctc pseudo labels,'' in \emph{ICASSP 2024 - 2024 IEEE International Conference on Acoustics, Speech and Signal Processing (ICASSP)}, 2024, pp. 11\,006--11\,010.

\bibitem{25}
J.~Johnson, M.~Douze, and H.~J{\'e}gou, ``Billion-scale similarity search with gpus,'' \emph{IEEE Transactions on Big Data}, vol.~7, no.~3, pp. 535--547, 2019.

\end{thebibliography}

\end{document}